# Design of Adiabatic MTJ-CMOS Hybrid Circuits

Fazel Sharifi, MZ Saif and Abdel-Hameed Badawy
Klipsch School of Electrical and Computer Engineering,
New Mexico State University, Las Cruses, USA
{fsharifi, mzsaif, badawy}@nmsu.edu

*Abstract*— **Low-power designs are a necessity with the increasing demand of portable devices which are battery operated. In many of such devices the operational speed is not as important as battery life. Logic-in-memory structures using nano-devices and adiabatic designs are two methods to reduce the static and dynamic power consumption respectively. Magnetic tunnel junction (MTJ) is an emerging technology which has many advantages when used in logic-in-memory structures in conjunction with CMOS. In this paper, we introduce a novel adiabatic hybrid MTJ/CMOS structure which is used to design AND/NAND, XOR/XNOR and 1-bit full adder circuits. We simulate the designs using HSPICE with 32nm CMOS technology and compared it with a non-adiabatic hybrid MTJ/CMOS circuits. The proposed adiabatic MTJ/CMOS full adder design has more than 7 times lower power consumtion compared to the previous MTJ/CMOS full adder.**

## I. Introduction

In the last decade, internet of thing (IoT) devices, portable electronics such as smartphones, tablets and sensors has increased dramatically. Most of these devices are battery operated and thus power consumption (battery-life) has become a critical design constraint. Therefore, researchers set out to discover new methods for designing low-power electronics [1]. A method for designing low power electronics that reduces the leakage power consumption is to use non-conventional CMOS devices and using emerging nanotechnologies. Some new emerging technologies are very appropriate to be utilized in low power applications [2]. Another method for reducing the dynamic power consumption is to recover the stored energy in the load capacitor instead of dissipating it as heat. This approach which operates based on energy recovery is known as adiabatic (reversible) circuit design [3].

The main problem with CMOS devices scaling down is the increase in leakage power and reduction in gate control [4]. Such that in nanoscale CMOS devices leakage power is an important component of the total energy consumption. Therefore, to continue chip density and performance scaling while maintaining low power, emerging devices and technologies such as quantum dot cellular automata (QCA), carbon nanotube field effect transistor (CNFET), single electron transistor (SET) and nano magnetic devices are attracting considerable attention as possible alternatives to CMOS devices [5-8]. Among these new technologies spin based devices have attracted attentions as a potential successor for CMOS because of its outstanding characteristics such as near-zero standby power, non-volatility, high integration density, etc. [9].

Moreover, ultra-low power architectures can be realized by logic-in-memory (LiM) paradigm, where memory elements are distributed over logic circuits [10-11]. Magnetic tunnel junction (MTJ) is a nonvolatile memory that has short access time, small dimensions and compatible with CMOS technology [12]. Therefore, it is most suited to use in logic-in-memory architectures. LiM structures using MTJs are very appropriate to low power designs, because the static power dissipation is almost zero in these circuits.

On the other hand, in modern integrated circuits and systems with high switching activity, dynamic power plays a significant role in power consumption. Dynamic power dissipation is due to charging and discharging of load capacitors during output switching. Adiabatic circuits are a family of circuits which reduces the dynamic power consumption by means of energy recovery to the supply voltage. In adiabatic circuits a multiphase clock controls the charging and discharging of the load capacitor. There are two types of adiabatic circuits, fully adiabatic and quasi adiabatic circuits. In fully adiabatic circuit the leakage power through the switches is the only power loss, while in quasi adiabatic circuits some non-adiabatic power losses exist [13,14].

In this paper, we present Spin-MTJ based nonvolatile adiabatic family of circuits. To the best of our knowledge, this study is the first on nonvolatile adiabatic circuits. The remainder of the paper is organized as follows: a brief review of MTJ devices and adiabatic circuits is presented in Section II. The proposed circuits are described and analyzed in Section III. Finally, Section IV concludes the paper.

## II. Background

### A. *Magnetic Tunnel Junction Reveiw*

Magnetic tunnel junction (MTJ) consists of two ferromagnetic (FM) layers (one of the layers is fixed and the other one is free) and an oxide barrier layer sandwiched between these two layers as shown in Figure 1. The oxide barrier has the ability to store data more than ten years which is verified by Time-Dependent Dielectric Breakdown (TDDB) experimental measurements [15,16].

Two possible configurations (parallel and antiparallel) can be materialized according to the FM layers alignment. Based on these two configurations, an MTJ shows low resistance (RP) or high resistance (RAP) characteristics [17]. Thus, we can use these characteristics to implement LiM designs.

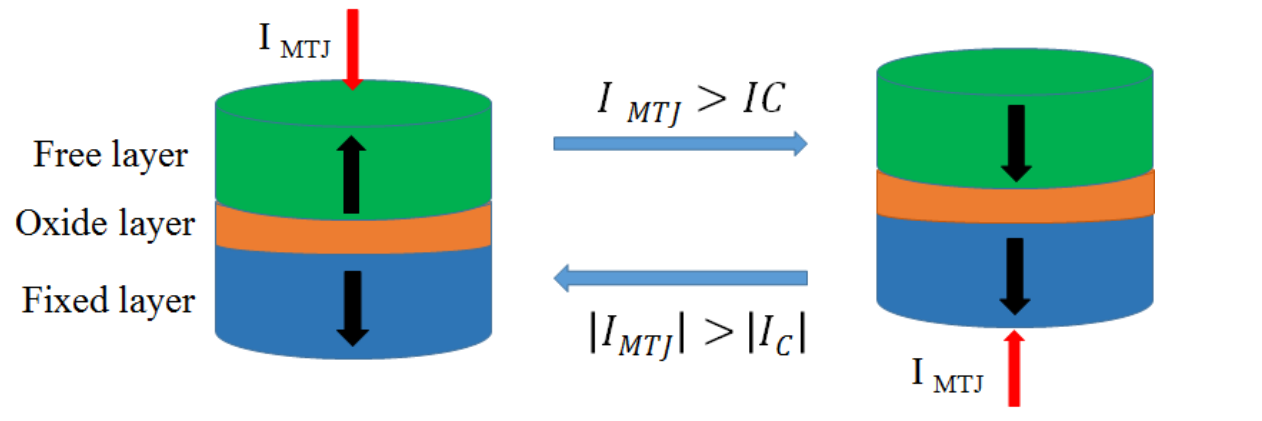

Fig. 1. Vertical Magnetic Tunnel Junction (MTJ) nanopillar structure. MTJ states change from P to AP and vice versa by applying a current ($I_{MTJ}$) higher than a critical current ($I_C$).

Three main methods have been proposed for switching MTJ configuration: Field Induced Magnetization Switching (FIMS), Thermally Assisted Switching (TAS) and Spin Torque Transfer (STT). The most promising method is STT which was proposed as an alternative for the other two methods. STT requires only one bi-directional low switching current. The states of the MTJ are switched when the current of MTJ ($I_{MTJ}$) becomes higher than a critical current ($I_C$) (Figure 1) [18]. FIMS was the conventional approach for switching MTJ states; RAP and RP, which were based on applying a magnetic field. This method suffers from high power consumption, poor selectivity, and poor scalability due to its high switching currents.

### B. Adiabatic Logic

Adiabatic logic is one of the low-power circuit design techniques at cost of slower speed of operation. The general schematic of an adiabatic technique is shown in Figure 2. In adiabatic circuits the load capacitance is charged by a constant current source unlike conventional CMOS where, the load capacitance is charged by a constant voltage source. Adiabatic logic reduces the overall power consumption of the circuit by employing a clocked AC power to charge the load capacitor and recovers energy from the charged capacitor in a slow manner to eliminate dynamic power dissipation [19].

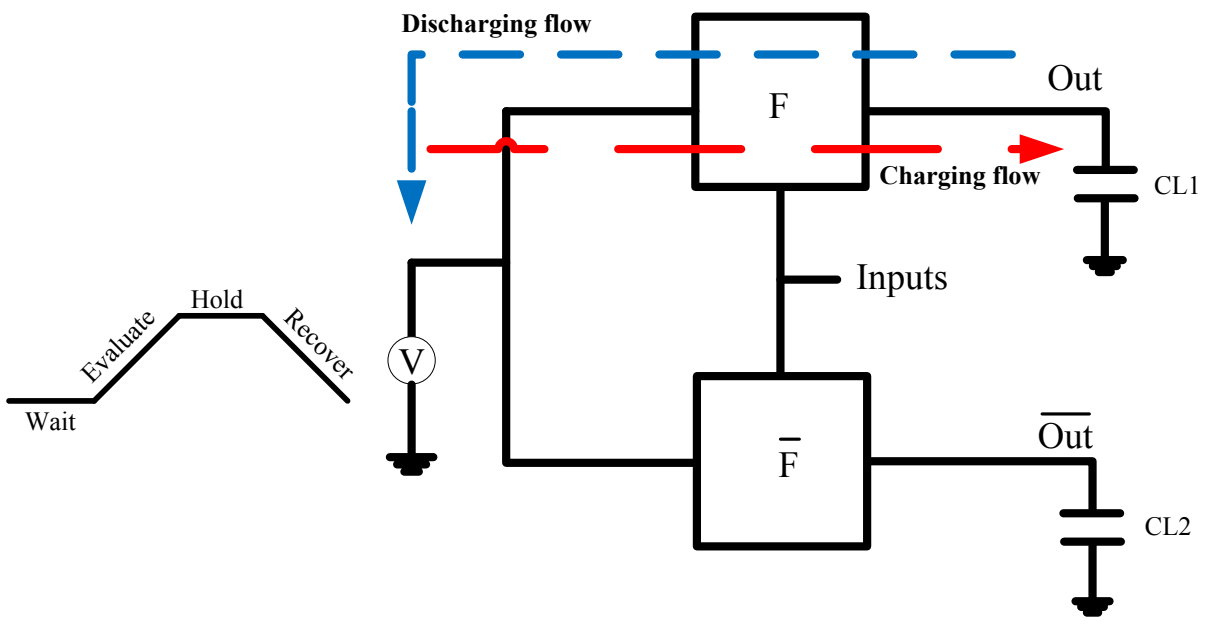

Fig. 2. Circuit representing adiabatic charging/discharging

The energy dissipated in an adiabatic circuit can be calculated based on the following equation [3]:

$$E_{diss} = \frac{RC}{T} C {V_{DD}}^2 \tag{1}$$

Where, C is the load capacitor, T is the capacitor charging time, and $V_{DD}$ is the full swing of the power clock.

In order to have less power consumption than conventional CMOS, adiabatic circuits have a charging time (T) that is greater than 2RC.

## III. Proposed Designs and Evaluation

In this section, we propose MTJ-based adiabatic circuits family. To the best of our knowledge this is the first attempt to design adiabatic magnetic circuits. MTJ-based circuits are generally composed of three parts as in Figure 3. The first part is a writing circuit, which is used for programing the memory elements. The second part constitutes of STT-MRAM cells and CMOS logic tree which as a logic control block. Finally, the last part is a sense amplifier (SA) that evaluates the output logic results.

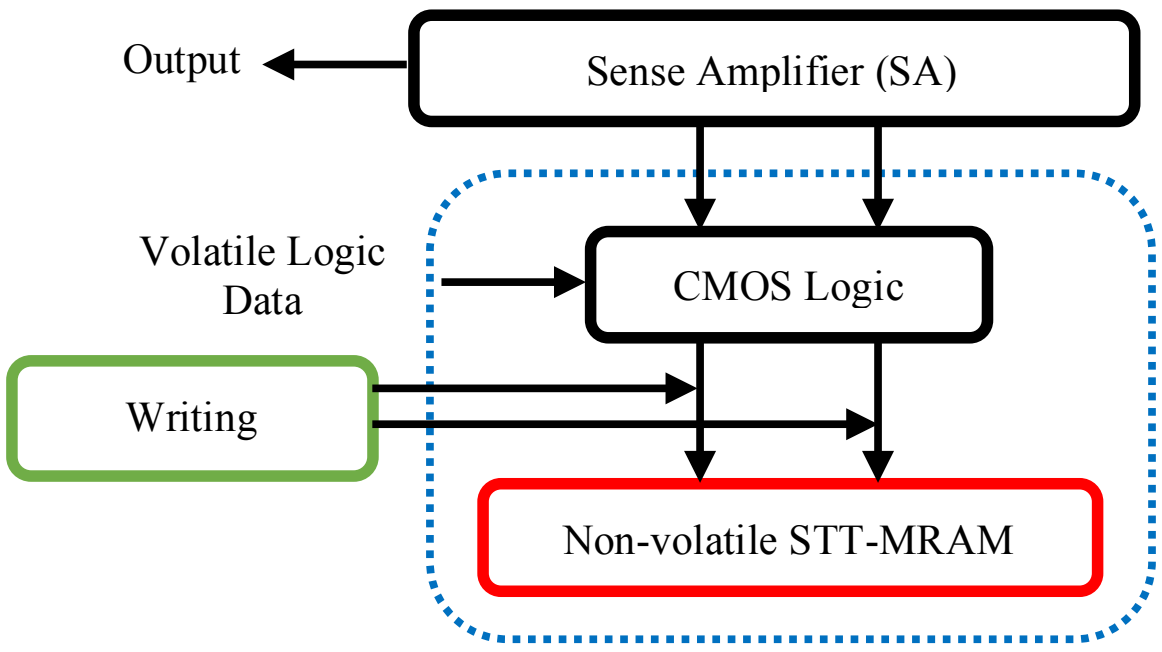

Fig. 3. Structure of a MTJ based circuit

A general view of the proposed adiabatic MTJ-based circuit family is in Figure 4. The difference between the presented circuits and MTJ-based circuits in the literature [12] is in the SA structure. To charge and discharge the outputs, a clocked AC power supply is used which has four phases (wait, evaluate, hold, and recover) as in Figure 2. Also, we use an N-MOS transistor to have an equal charge in the wait phase. Thus, the discharge signal is $V_{DD}$ when the circuit is in the wait phase.

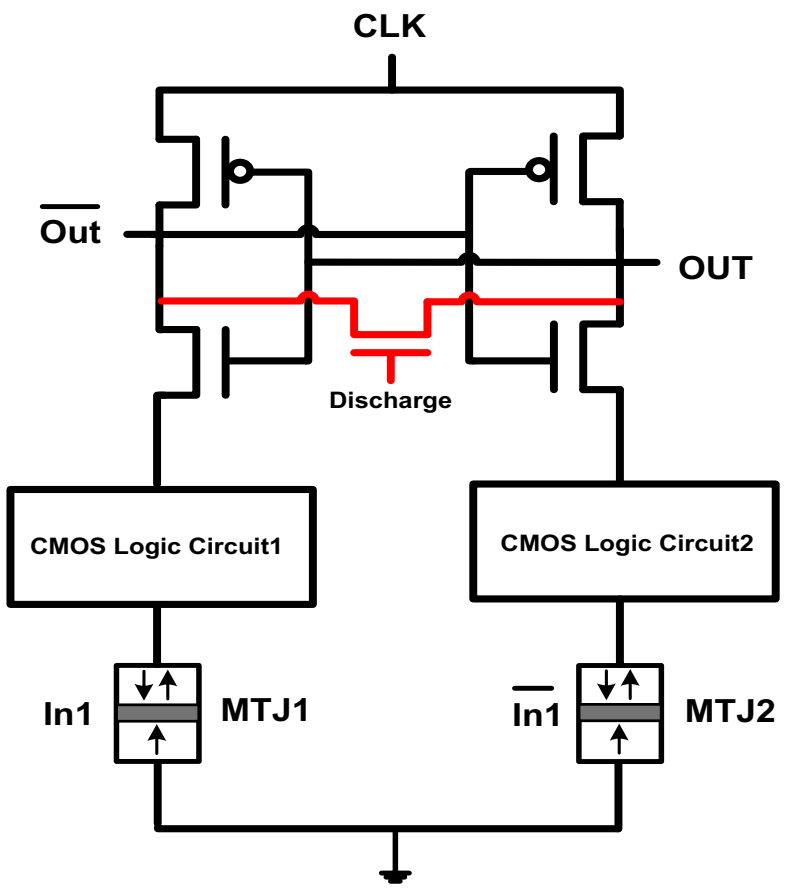

Fig. 4. General schematic of the proposed adiabatic hybrid MTJ/CMOS design. The initial states of MTJ and MTJ2 are antiparallel and parallel respectively.

In the proposed method, we eliminated two P-FETs and an N-FET from the circuits presented in [12].

In this paper, adiabatic MTJ-based logic gates and arithmetic circuits including AND/NAND, XOR/XNOR gates and a full adder cell are presented in the next subsections.

### A. *Logic Gates (AND/NAND, XOR/XNOR)*

The schematic design of the AND gate is shown in Figure 5. When the CLK is in the wait phase, both AND as well as NAND outputs are zero. Assume that the input pattern is "01" for "AB" then, T5 will be off and MTJ1 and MTJ2 will be in parallel and antiparallel states, respectively. With this input pattern, the left path is cut off and the AND output will be discharged to the ground and consequently the NAND output will charged to $V_{DD}$ in the evaluate phase. The outputs will remain the same in the hold phase whereas in the recovery phase, the NAND output will be discharged to the CLK supply power. Since T1 and T2 cannot discharge the outputs completely to zero, in the next wait phase the discharge signal will be $V_{DD}$ to share the outputs voltages and both outputs have the same amount of voltage.

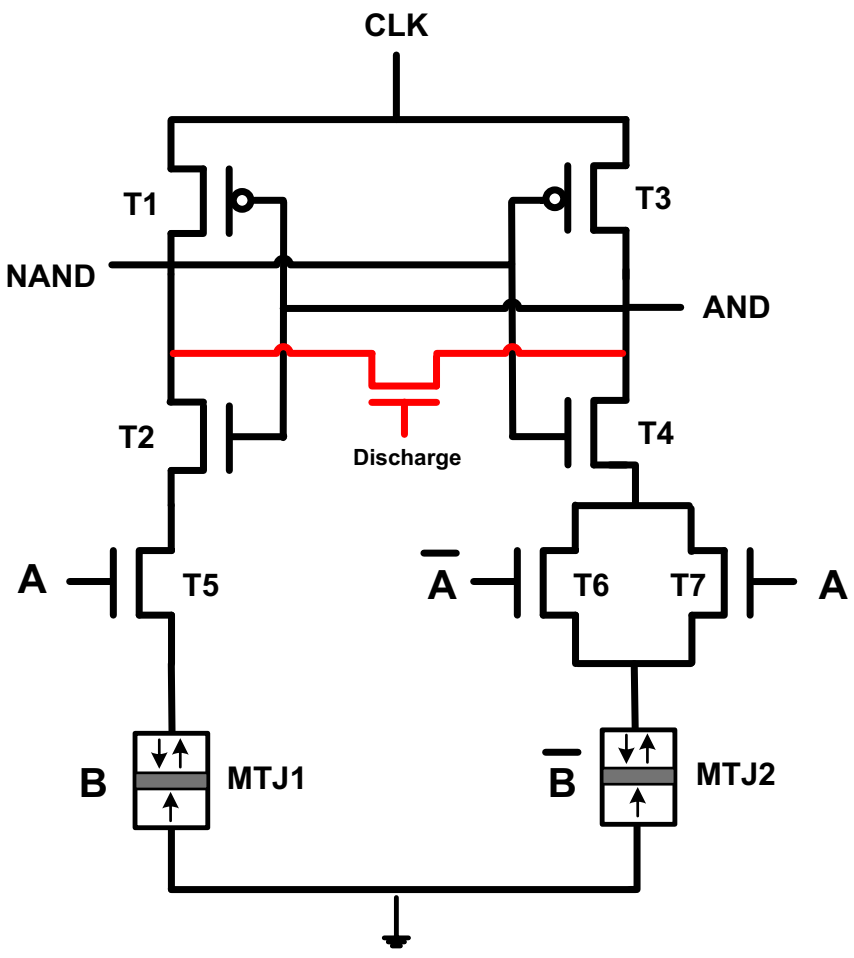

Fig. 5. Proposed adiabatic hybrid MTJ/CMOS AND/NAND.

In the following we elaborate more on the functionality of the proposed XOR as in Figure 6.

Consider that the input pattern is "00" for "AB" and the initial states of MTJ1 and MTJ2 are antiparallel and parallel, respectively. At the wait phase, the memories are programed since the voltage level of the clock is zero. Thus, the states of MTJ1 and MTJ2 remain at the previous states because the voltage of input B was not changed. At the evaluate phase the writing circuit is disconnected from the main circuit and the voltage level starts to increase. With this input pattern, T2 and T3 are ON and MTJ2 has a lower resistance than MTJ1. Since the left path has more resistance (because of the antiparallel state of MTJ1), the XNOR output will be charged and the XOR output will remain zero in the Evaluation phase. The voltage level of the XOR and XNOR outputs will remain at 0 and VDD in the hold phase. Finally, in the recovery phase the outputs will be discharged and recovered to the CLK. Now, consider the input patterns go to "01" from "00" in the wait Phase, so the states of MTJ1 and MTJ2 will change from antiparallel to parallel and vice versa, respectively. T2 and T3 will be ON and MTJ1 will have lower resistance than MTJ2. Accordingly, the resistance value of the right path will be higher than the left path, and the XOR output will be charged during the evaluate phase. When the inputs value is "11", the states of MTJs will remain the same. In this pattern, T1 and T4 are ON and since MTJ2 has more resistance, the XOR output will remain 0V and XNOR will be charged in the evaluate phase. During the hold phase, the outputs will not be changed and in the recovery phase the XOR output will follow the CLK voltage to zero.

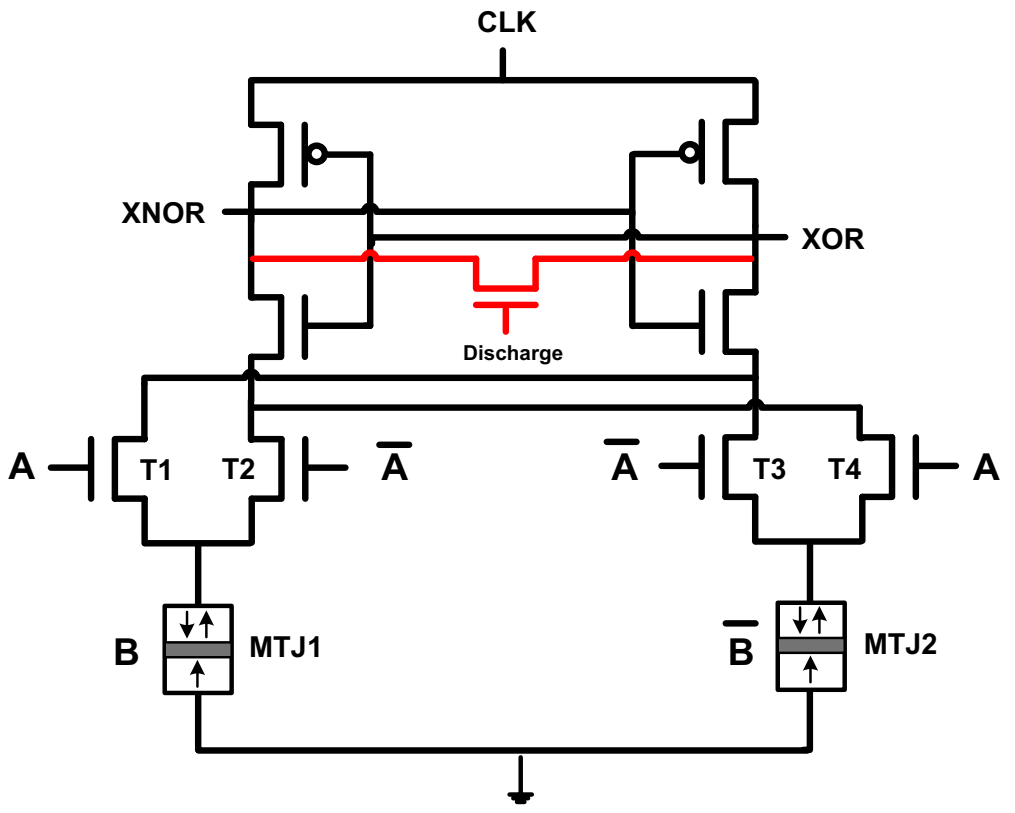

Fig. 6. Proposed adiabatic hybrid MTJ/CMOS XOR/XNOR.

### B. *Full Adder Cell*

Figure 7 depicts a schematic of the proposed full adder cell. The initial state of the MTJ1 and MTJ3 is antiparallel and the initial state of the MTJ2 and MTJ4 is parallel. The principle of operation of this circuit is similar to the proposed XOR circuit but with a different output. The Sum and Carry out of the adder are given in equations 2--5.

$$Sum = ABC + A\bar{B}\bar{C} + \bar{A}B\bar{C} + \bar{A}\bar{B}C \quad (2)$$

$$\overline{Sum} = AB\bar{C} + A\bar{B}C + \bar{A}BC + \bar{A}\bar{B}\bar{C} \quad (3)$$

$$Cout = AB + AC + AC \quad (4)$$

$$\overline{Cout} = \bar{A}\bar{B} + \bar{A}\bar{C} + \bar{B}\bar{C} \quad (5)$$

For example, when the inputs pattern is "001" for "ABC", the $\overline{Sum}$ output will be zero via the T2, T8 and MTJ2 and consequently the Sum will be charged in the evaluate phase. Also, Cout will be zero through the path of T11 and MTJ4.

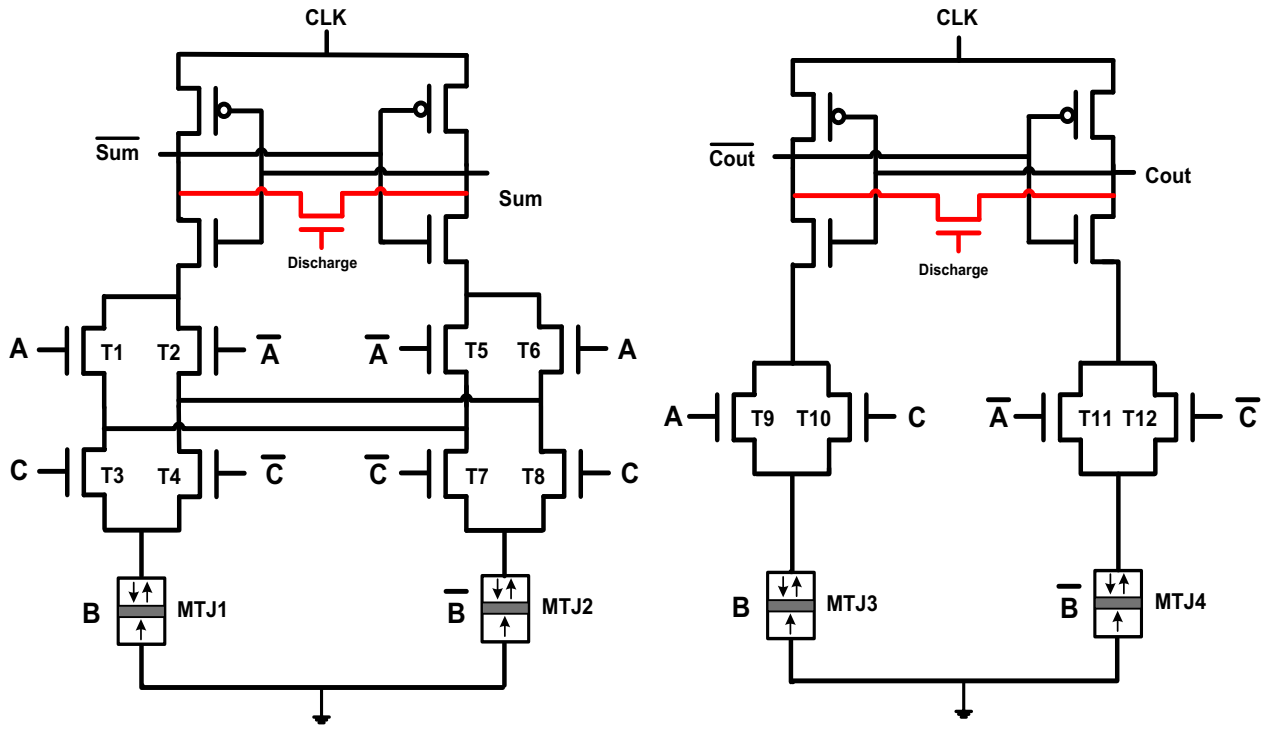

Fig. 7. Proposed adiabatic hybrid MTJ/CMOS full adder design.

### C. *Designs Evaluation and Comparison*

The proposed designs are simulated and compared with the MTJ/CMOS design in [12] in terms of power consumption. Simulations are conducted using the HSPICE circuit simulator with 32nm technology for CMOS transistors [20] and the spice MTJ model presented in [21] for MTJ devices. Figure 8 shows the transient response of the proposed XOR design when the input B is $V_{DD}$. It confirms the correct operation of our design. The comparison results of our proposed designs and the design in [12] are depicted in Figure 9. The graphs suggest that the proposed adiabatic hybrid MTJ/CMOS XOR, AND, and the full adder designs have almost 13, 6 and 7 times lower power consumption compared to the designs presented in [12].

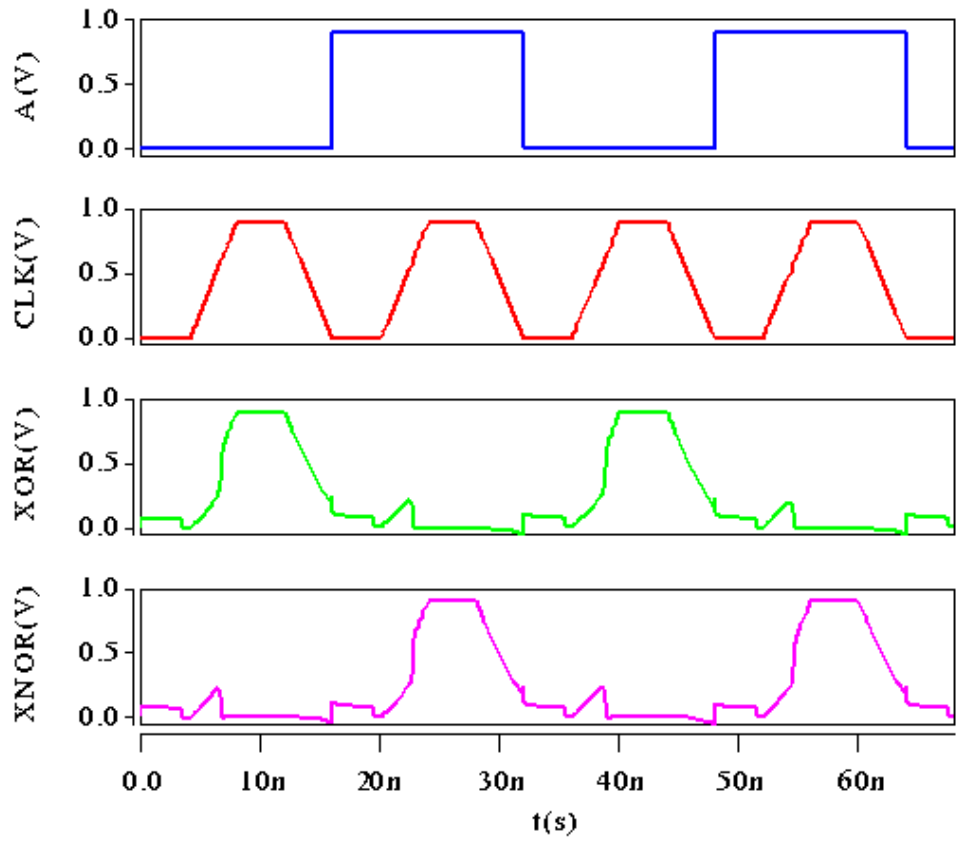

Fig. 8. Transient response of the proposed adiabatic hybrid MTJ/CMOS full adder circuit. Here it is assumed that the input "B" is logic "1" and stored in MTJs.

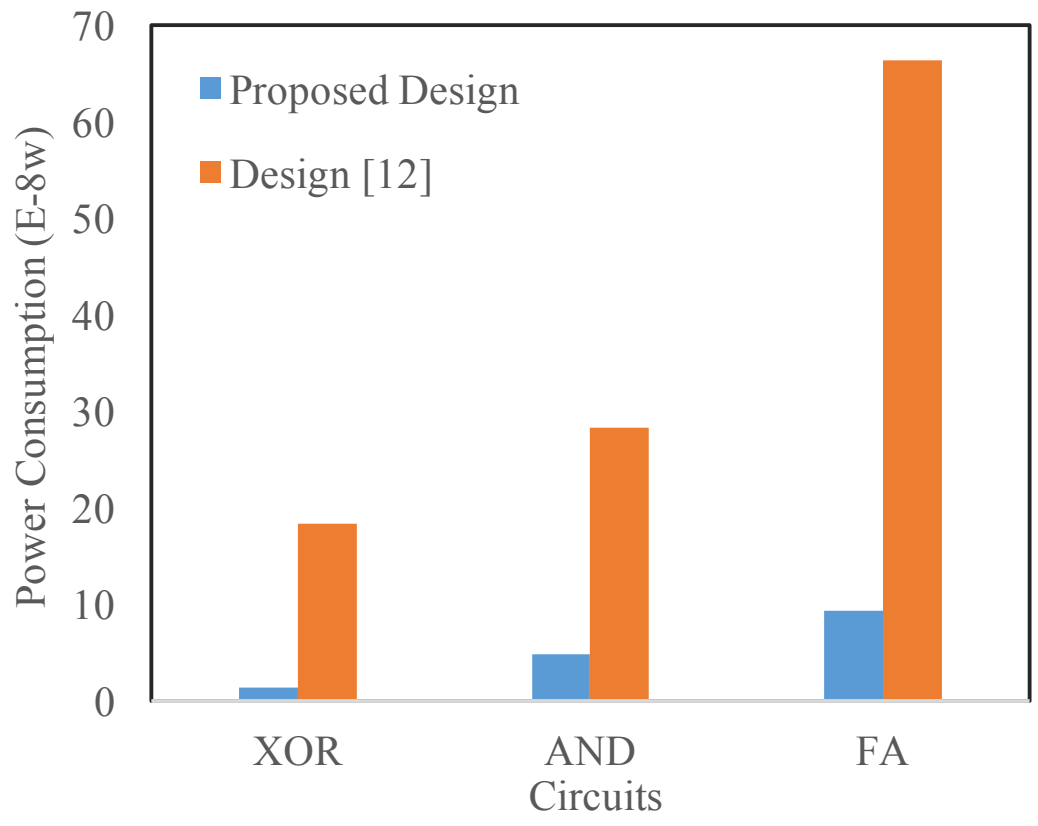

Fig. 9. Power consumption of the proposed designs and designs [12].

## IV. CONCLUSION

Logic-in-Memory (LiM) structures that use magnetic devices and adiabatic designs are two efficient approaches to realize low power designs. A new structure for designing adiabatic hybrid MTJ-CMOS circuits is presented in this paper. We have implemented AND/NAND, XOR/XNOR, and full adder circuits with this structure. Designs are simulated and compared with state of the art. We used Synopsys HSPICE simulator with 32 nm technology files to evaluate our designs. The results show that the proposed adiabatic MTJ-CMOS designs have lower power consumption compared to state of the art, such that the proposed XOR, AND, and full adder have almost 13, 6, and 7 times lower power consumption respectively compared to state of the art.

## REFERENCES

[1] M. Pedram and J. M. Rabaey, Power aware design methodologies: Springer Science & Business Media, 2002.

[2] Y. Zhang, W. Zhao, J.-O. Klein, W. Kang, D. Querlioz, Y. Zhang, et al., "Spintronics for low-power computing," in 2014 Design, Automation & Test in Europe Conference & Exhibition (DATE), 2014, pp. 1-6.

[3] [3] W. C. Athas, L. J. Svensson, J. G. Koller, N. Tzartzanis, and E. Y.-C. Chou, "Low-power digital systems based on adiabatic-switching principles," IEEE Transactions on Very Large Scale Integration (VLSI) Systems, vol. 2, pp. 398-407, 1994.

[4] N. S. Kim, T. Austin, D. Baauw, T. Mudge, K. Flautner, J. S. Hu, et al., "Leakage current: Moore's law meets static power," computer, vol. 36, pp. 68-75, 2003.

[5] I. Amlani, A. O. Orlov, G. Toth, G. H. Bernstein, C. S. Lent, and G. L. Snider, "Digital logic gate using quantum-dot cellular automata," science, vol. 284, pp. 289-291, 1999.

[6] S. Lin, Y.-B. Kim, and F. Lombardi, "CNTFET-based design of ternary logic gates and arithmetic circuits," IEEE transactions on nanotechnology, vol. 10, pp. 217-225, 2011.

[7] M. H. Sulieman and V. Beiu, "On single-electron technology full adders," IEEE Transactions on Nanotechnology, vol. 4, pp. 669-680, 2005.

[8] H. Lee, F. Ebrahimi, P. K. Amiri, and K. L. Wang, "Low-Power, High-Density Spintronic Programmable Logic With Voltage-Gated Spin Hall Effect in Magnetic Tunnel Junctions," IEEE Magnetics Letters, vol. 7, pp. 1-5, 2016.

[9] J. Kim, A. Paul, P. A. Crowell, S. J. Koester, S. S. Sapatnekar, J.-P. Wang, et al., "Spin-based computing: device concepts, current status, and a case study on a high-performance microprocessor," Proceedings of the IEEE, vol. 103, pp. 106-130, 2015.

[10] S. Matsunaga, J. Hayakawa, S. Ikeda, K. Miura, H. Hasegawa, T. Endoh, et al., "Fabrication of a nonvolatile full adder based on logic-in-memory architecture using magnetic tunnel junctions," Applied Physics Express, vol. 1, p. 091301, 2008.

[11] W. H. Kautz, "Cellular logic-in-memory arrays," IEEE Transactions on Computers, vol. 100, pp. 719-727, 1969.

[12] E. Deng, Y. Zhang, J.-O. Klein, D. Ravelsona, C. Chappert, and W. Zhao, "Low power magnetic full-adder based on spin transfer torque MRAM," IEEE transactions on magnetics, vol. 49, pp. 4982-4987, 2013.

[13] J. Hu, T. Xu, J. Yu, and Y. Xia, "Low power dual transmission gate adiabatic logic circuits and design of SRAM," in Circuits and Systems, 2004. MWSCAS'04. The 2004 47th Midwest Symposium on, 2004, pp. I-565.

[14] J. Lim, D.-G. Kim, and S.-I. Chae, "nMOS reversible energy recovery logic for ultra-low-energy applications," IEEE Journal of Solid-State Circuits, vol. 35, pp. 865-875, 2000.

[15] J. S. Moodera, L. R. Kinder, T. M. Wong, and R. Meservey, "Large magnetoresistance at room temperature in ferromagnetic thin film tunnel junctions," Physical review letters, vol. 74, p. 3273, 1995.

[16] R. Zand, A. Roohi, S. Salehi, and R. DeMara, "Scalable Adaptive Spintronic Reconfigurable Logic using Area-Matched MTJ Design."

[17] B. Behin-Aein, J.-P. Wang, and R. Wiesendanger, "Computing with spins and magnets," MRS Bulletin, vol. 39, pp. 696-702, 2014.

[18] W. Zhao, E. Belhaire, C. Chappert, and P. Mazoyer, "Spin transfer torque (STT)-MRAM--based runtime reconfiguration FPGA circuit," ACM Transactions on Embedded Computing Systems (TECS), vol. 9, p. 14, 2009.

[19] R. K. Yadav, A. K. Rana, S. Chauhan, D. Ranka, and K. Yadav, "Adiabatic technique for energy efficient logic circuits design," in Emerging Trends in Electrical and Computer Technology (ICETECT), 2011 International Conference on, 2011, pp. 776-780.

[20] http://ptm.asu.edu/

[21] J. Kim, A. Chen, B. Behin-Aein, S. Kumar, J.-P. Wang, and C. H. Kim, "A technology-agnostic MTJ SPICE model with user-defined dimensions for STT-MRAM scalability studies," in *Custom Integrated Circuits Conference (CICC), 2015 IEEE*, 2015, pp. 1-4.